\documentstyle[agums]{article}
\begin{document}

\author{R. E. Cohen and J. S. Weitz \\
%EndAName
Geophysical Laboratory and Center for High Pressure Research, Carnegie
Institution of Washington, Washington D.C. 20015}
\title{The Melting Curve and Premelting of MgO}
\date{\today}

\begin{abstract}
The melting curve for MgO was obtained using molecular dynamics and a
non-empirical, many-body potential. We also studied premelting effects by
computing the dynamical structure factor in the crystal on approach to
melting. The melting curve simulations were performed with periodic boundary
conditions with cells up to 512 atoms using the ab-initio Variational
Induced Breathing (VIB) model. The melting curve was obtained by computing $%
\Delta H_m$ and $\Delta V_m$ and integrating the Clapeyron equation. Our $%
\Delta H_m$ is in agreement with previous estimates and we obtain a
reasonable $\Delta V_m$, but our melting slope dT/dP (114 K/GPa) is three
times greater than that of Zerr and Boehler [1994] (35 K/GPa), suggesting a
problem with the experimental melting curve, or an indication of exotic,
non-ionic behavior of MgO liquid. We computed $S(q,\omega )$ from
simulations of 1000 atom clusters using the Potential Induced Breathing
(PIB) model. A low frequency peak in the dynamical structure factor $%
S(q,\omega )$ arises below the melting point which appears to be related to
the onset of bulk many-atom diffusive exchanges. These exchanges may help
destabilize the crystalline state and be related to intrinsic crystalline
instability suggested in earlier simulations.
\end{abstract}

\section{Introduction}

Understanding melting is crucial for understanding the evolution and
dynamics of the Earth. In order to trace the development of the Earth from
its origin until now, it is important to know the melting temperatures,
enthalpy of melting, and density of melts and solids as functions of
composition. There is also fundamental interest in understanding the melting
process. Why and how do crystals melt, and are there any precursors to the
melting transition evident in the crystalline phase? Here we are
particularly interested in how pressure might effect melting and premelting
behavior. Since we are interested in eventually understanding melting in the
Earth, we start with the simplest oxide, MgO, and study its melting and
premelting in the crystalline phase as a function of compression.

Measuring melting curves to extreme pressures is very difficult, and there
have been significant discrepancies among laboratories on melting curves of
the important geophysical materials Fe [{\it Anderson and Ahrens}, 1996; 
{\it Boehler}, 1996] and MgSiO$_3$ perovskite [{\it Heinz et al.}, 1994; 27; 
{\it Sweeney and Heinz}, 1993; {\it Zerr and Boehler}, 1993]. It is also
very difficult to calculate melting curves theoretically in spite of many
attempts to develop predictive models for melting. Calculation of melting
curves from fundamental physics is a difficult undertaking as well;
difficult because accurate potentials or electronic structure methods are
needed to obtain the forces among atoms, long simulations are needed to
equilibrate and obtain thermodynamic properties of the liquid, and since
free-energies cannot be directly calculated, one must perform thermodynamic
integrations or reversals [{\it Cohen and Gong}, 1994] in order to obtain
the melting point. The most widely applied method to obtain melting points
presently is by thermodynamic integration, where one starts with some
reference system such as the ideal gas where the free energy is known, and
then slowly varies the temperature and pressure until the desired liquid
state at P and T is reached. This would be difficult for ionic systems,
however, where there is no ergodic ideal reference state for the charged
system. [One could use the ionic charge itself as the integration parameter, and slowly vary the charge until the final charged state is reached.
However, this would require many simulations to integrate accurately from a
reference state of such different character than the final state.]

Cohen and Gong [1994] predicted the melting curve of MgO to 300 GPa
using molecular dynamics simulations for finite clusters. The
interactions among atoms were obtained using the non-empirical
Potential Induced Breathing (PIB) model which had been shown
previously to give excellent agreement with experiment for
thermoelastic properties of MgO to high pressures [{\it Isaak,
Cohen, and Mehl}, 1990]. Molecular dynamics simulations using the
closely related VIB (Variationally Induced Breathing) model
(described below) show exceptional accuracy for the equation of
state of MgO, including high order properties such as the change in
thermal expansivity with pressure [{\it Inbar and Cohen},
1995]. Given the accuracy of the model for properties of the solid,
it was surprising when the first experimental measurements for the
melting curve of MgO showed a discrepancy of over a factor of three
in the dT/dP slope, with the experiments showing a much shallower
slope [{\it Zerr and Boehler}, 1994]. Since the MD and lattice
dynamics calculations showed that crystalline properties of MgO were
well predicted by the models, such a large discrepancy could
indicate a problem with the liquid simulations. Since the potentials
do not use any information that is particular to the crystalline
state and are based on fundamental physics, only one possibility
seemed open--that there was a problem with the liquid simulations
due to the use of finite clusters. Cohen and Gong [1994] used finite
clusters of 64 to 1000 atoms and then extrapolated to the bulk; they
found that the finite cluster results were linear functions of $1/L$
, where $L$ is the length (i.e. linear dimension) of the clusters, so they
extrapolated T$_m$ with $1/L\rightarrow 0$. The extrapolation to bulk
is effectively over about 20 orders of magnitude of system size, so
that there is cause for concern that this could introduce errors in
the predicted melting curve. Thus here we have effectively
eliminated size effects by using another technique based on similar
potentials, but with periodic boundary conditions and no surfaces,
to obtain the melting slope.

We also further continue the study of melting in clusters using PIB, and
look for dynamical premelting effects by computing and studying the power
spectrum S(q,$\omega $) in the crystalline phase. Cohen and Gong [1994]
found evidence of an intrinsic instability in the crystal near melting by
studying the Lindemann ratio $u_{rms}/a$ where $u_{rms}$ is the r.m.s.
displacement and $a$ is the mean near neighbor distance. They found this
ratio to be constant along the melting curve spanning 300 GPa and 15,000 K.
Whereas such scaling is expected in power law potentials [{\it Ross}, 1969]
where liquid and solid structures are constant along the melting curve, it
is not constrained to behave thusly with realistic potentials. The large
changes in liquid structure along the melting curve indicate that the
constancy of the Lindemann ratio must have a deeper origin. Cohen and Gong
[1994] hypothesized that in pure systems such as MgO, there might be an
underlying instability in the crystal leading to melting. The fact that
melting is a first-order transition in no way negates this possibility; for
instance, there is universal agreement that there is an
underlying and observable overdamped soft mode in BaTiO$_3$, in spite
of the fact that its ferroelectric transitions  are first-order. They
proposed a similar scenario for MgO, but there was no evidence of what the
underlying instability might be. The best candidate is the shear instability
c$_{11}$-c$_{12}$, which vanishes at the melting point at zero pressure, but
the instability occurs at higher temperatures than the melting point with
increasing pressure; the meaning of this behavior is still a mystery. Here
we try to understand better whether there is an underlying dynamical
instability in the lattice and what it is by studying the dynamical
structure factor or power spectrum S(q,$\omega $) in crystalline clusters of
1000 atoms as melting is approached.

Much work has been done on premelting in clusters and finite systems
concentrating on surface melting and roughening transitions [{\it %
Bastiannsen and Knops}, 1996; {\it Nagaev and Zil'bervarg},
1996]. There is also considerable evidence that small clusters do
not undergo discontinuous phase transitions, but rather go through a
region of fluctuations between solid-like and liquid-like
configurations [{\it Bhattacharya, Chen, and Mahanti}, 1996; {\it
Nayak, Ramaswamy, and Chakravarty}, 1995; {\it Wells and Berry},
1994]. Our 1000 atom clusters are significantly larger than those
discussed in the latter studies, which range up to 55 atoms. Evidence
was seen in Cohen and Gong [1994] for coexistence and slow
fluctuations between solid and liquid in the van der Waals loop
region of the transition, which is probably closely related to what
is observed in the smaller clusters. Our interest in studying
S(q,$\omega $) is not however to understand better the dynamics of
this fluctuation/coexistence regime, but rather to look for evidence
of an approaching dynamical instability in the crystalline field as
the melting point is approached.

\section{Methods}

We performed classical molecular dynamics simulations for periodic and
cluster systems, using the non-empirical VIB and PIB models, respectively.
VIB [{\it Wolf and Bukowinski}, 1988] and PIB [{\it Cohen, Boyer, and Mehl},
1987a] are very similar ionic Gordon-Kim [{\it Gordon and Kim}, 1972] type
models, in which the total charge density is modeled by overlapping ionic
charge densities, which are computed from quantum mechanical atomic
calculations with no adjustable parameters. Only the ionic charge and
nuclear charge are input, and the ionic charges used are the nominal 2+ and
2- for Mg and O, respectively. In Gordon-Kim models, the total energy is a
sum of three terms, the long-range electrostatic or Madelung energy, the
self-energy of each atom or ion, and the short-range interaction energy
which is a sum of the kinetic energy, short-range electrostatic, and
exchange-correlation energy, all of which are functions of the model charge
density. We use the Hedin-Lundqvist [{\it Hedin and Lundqvist}, 1971]
parametrization of the exchange-correlation energy and the Thomas-Fermi
kinetic energy in the interaction energy. For the self-energies we use the
Kohn-Sham total energy [{\it Kohn and Sham}, 1965].

Since O$^{2-}$ is not stable in the free state, it is stabilized with a
sphere of +2 charge (called a ``Watson sphere'') in the atomic calculations,
and the radius of this sphere is chosen different in the PIB and VIB models;
this is the only difference between PIB and VIB. In the PIB\ model, the
radius of the Watson sphere, $r_{Wat}$ is chosen so that the electrostatic
potential inside the sphere is the same as the Madelung (i.e.\thinspace
electrostatic) potential at the site the crystal in order to model the
electrostatic stabilization of the O$^{2-}$ ion by the crystal field. In the
VIB model, $r_{Wat}$ is chosen to minimize the total energy in the crystal
for a given configuration of atoms. Both VIB and PIB give very similar
results except at very high pressures where the VIB potential is softer and
more accurate due to the fact that it includes short-range contributions to
the O$^{2-}$ size, as well as the long-range Madelung contributions [{\it %
Inbar and Cohen}, 1995]. They also give different LO-TO splittings; VIB
gives the rigid ion LO-TO splitting whereas PIB gives reduced values that
are closer to experiment (although for the wrong reasons) when a simple
correction to reference the Madelung potential to the local average
potential is included. This is due to the dependence in PIB of $r_{Wat}$ on
the magnitude of the potential, as opposed to potential differences [{\it %
Cohen, Boyer, and Mehl}, 1987a, 1987b].  Since the VIB model is better
behaved we used the VIB choice of $r_{Wat}$ here for the periodic boundary
condition computations. We employ the pair approximation, and calculate
pairwise interactions as functions of the distance between atoms and $r_{Wat}
$ on the anions, and fit an analytic function to the calculated energies as
functions of distance and $r_{Wat}$. This function, which has up to 21
parameters for O-O interactions, can be evaluated much more rapidly than
doing the full quantum calculations at each time step. The resulting
potential has been severely tested for thermal properties of MgO
[{\it Isaak et al., 1990; Inbar and Cohen, 1995}], and we
have great confidence in the potential.

In the molecular dynamics simulations, Newton's equation ${\bf F}={\bf ma}$
is integrated forward in time. In VIB, the energy is minimized with respect
to all of the ${\bf r}_{Wat}$ for each time step. In PIB, the Madelung
potentials are computed at each time step. In both cases, the forces are
obtained analytically for each configuration of atoms and Watson sphere
radii $r_{Wat}$. The pair interactions and the self-energies of the O$^{2-}$
anions are functions of $r_{Wat}$ for each anion.

\subsection{Melting Curve}

We have simulated crystalline and liquid MgO with periodic boundary
conditions, i.e.\thinspace with no surfaces, using the VIB model to obtain
the melting curve. Periodic boundary conditions introduce long-range lattice
structure onto a liquid, which should not be present, and this can cause
systematic errors especially in ionic crystals with long-range forces.
However, by studying two very different periodic cell sizes, 64 atoms, and
512 atoms, we can test whether the quantities we calculate, V and T for
given P and E, are affected. We obtain the change in enthalpy and volume, $%
\Delta H=\Delta E+P\Delta V$ and $\Delta V$, as functions of T and P between
the solid and liquid, which at the melting point$\ T_m$ gives us the melting
slope through the Clapeyron equation, 
\begin{equation}
dT/dP=\frac{T_m\Delta V_m}{\Delta H_m}.
\end{equation}
Since the primary discrepancy with experiment is the melting slope, and both
theory and experiment agree on the melting point at zero pressure, we fix
the zero pressure melting point at 3200K and then integrate the Clapeyron
equation to give the melting curve.

In the periodic boundary condition simulations, we employed the
variable-cell-shape technique of Parinello and Rahman[1980], in which extra
fictitious dynamical degrees of freedom associated with the shape of the
computational cell are introduced to allow the computational cell volume and
shape to vary. This technique conserves enthalpy, rather than energy, and
the external pressure is kept constant, rather than the volume. The
off-diagonal elements of the strain matrix were not allowed to vary to avoid
problems with large fluctuations in the shape of the periodic cell in the
liquid state.

The systems consisted of a sample of 64 atoms, initially arranged in
a cubic lattice. In order to check for system-size effects, some
simulations were also carried out on a 512- atom system. Periodic
boundary conditions were employed to eliminate surface effects, and
a timestep of 1 fs was selected.  The equations of motion were
numerically integrated using a fifth-order Gear Predictor-Corrector
method [{\it Gear, 1966}]. Throughout our simulations, enthalpy was
conserved to approximately 1 part in 10$^6$ per iteration.

Simulations were performed at P= 0, 12.5, 25, 50 and 100 GPa. At each
pressure, MD runs were performed at various temperatures near the expected
melting point in both solid and liquid. Initially, the kinetic energies of
the atoms were scaled to obtain approximately the desired temperature. After
equilibration, which lasted for 2 ps, we ran each simulation for an
additional 6-15 ps during which the system volume, enthalpy, and kinetic
energy were monitored each iteration. The length of each run was determined
by the convergence of the average volume of the system. The enthalpy was a
constant of the motion. The temperature was calculated from the average
kinetic energy, and the volume averaged.

After performing these simulations at several temperatures in the solid, the
limit of superheating was reached, and the solid melted. This was determined
both from the presence of diffusion, and the dramatic reduction in intensity
of a simulated Bragg reflection intensity. After melting, the temperature
dropped because of the conversion of the latent heat of melting to potential
energy. Similar simulations were performed in the liquid.

\subsection{Power Spectrum}

Since the previous cluster calculations [{\it Cohen and Gong}, 1994] used
PIB as opposed to VIB, we use PIB here for the study of the dynamic
structure factor S(q,$\omega $) in clusters. The power spectrum S(q,$\omega $%
) was obtained as follows. Simulations for clusters of 1000 atoms were
performed for MgO at P=0, 100 and 150 GPa at a series of temperatures. At
zero pressure the cluster had free boundary conditions, and for the high
pressure runs pressure was imposed by enclosing the cluster in a cubic
elastic box as in Cohen and Gong. Atoms that hit the box walls reflect
specularly; the momentum component perpendicular to the wall is reversed.
The Verlet algorithm was used to integrate the classical Newton's equations
with a time step of 2 fs. Simulations were started with equilibrated system
prepared by Cohen and Gong [1994] and were run for 20,000 time steps. Frames
of the atomic positions were saved every 8 time steps for computation of S(q,%
$\omega $). The density function $\rho $ is defined as

\begin{equation}
\rho ({\bf r},t)=\sum_{i=1}^N\delta ({\bf r}-{\bf r}_i(t))
\end{equation}
and the transform is

\begin{equation}
p({\bf k},t)=\sum_{i=1}^Ne^{i{\bf k}\cdot {\bf r}_i(t)}.  \label{density}
\end{equation}
The dynamical structure factor is defined as

\begin{equation}
S({\bf q},\omega )=\int dtF({\bf q},t)e^{i\omega t}W(t)  \label{S}
\end{equation}
where ${\bf q}=2\pi {\bf k/l}$ for mean cubic cell lattice constant $l$ and $%
F({\bf q},t)$ is the intermediate structure function defined as the
correlation function

\begin{equation}
F({\bf q},t)=\left\langle p({\bf q},t)p^{*}({\bf q},0)\right\rangle
\end{equation}
where $\left\langle {}\right\rangle $ indicates an average over all time
origins. We used the Blackman-Harris exact three parameter window function [{\it %
Harris}, 1978]

\begin{equation}
W(t)=0.42659071+0.49656062\cos \left( 2\pi t/T\right) +0.07684867\cos \left( 4\pi t/T\right) 
\end{equation}
where T is the total time in order to minimize artifacts from the finite
time series. Eq. \ref{S} is solved by a discrete fast Fourier transform over
the time slices. In order to improve statistics [{\it Press et al.}, 1992]
runs were divided into eight segments and seven overlapping transforms were
performed and averaged. We also averaged over equivalent $q$'s and the final 
$S({\bf q},\omega )$ was smoothed with a running average over $\pm 1$ cm$%
^{-1}$.

There are non-trivial issues regarding {\bf q} dependent quantities such as $%
S({\bf q},\omega )$ for clusters. In periodic boundary conditions, the
meaningful {\bf q}-space is quantized according to the size of the periodic
cell. However in a cluster, results can be obtained for any q. This is
because we have free boundary conditions at the surfaces so that the waves
do not need to have nodes at the surfaces. However for almost all q's, there
is a large peak at $S(q,\omega =0)$ due to the fact that there are not
equal numbers of positive and negative displacements for most q's. At these
q's it is difficult to obtain a clear spectrum of $S(q,\omega )$ because
aliasing and spillout of the $\omega=0$ peak results from the finite time
sampling and windowing, thus masking the physically important behavior at
small $\omega$. Thus we pick q's that are minima in $S(q,0)$, and these turn
out to be close to the commensurate q's at $\left( \frac q2\frac q2\frac
q2\right) $, $\left( qq0\right) $, and $\left( q00\right) $ where $%
q=0.1,0.2,0.3,0.4,$and $0.5$ $4\pi /a$ where $a$ is the cubic lattice
constant (note that for an fcc lattice the X point is at $(2\pi /a,0,0)$ and
the L point is at $(\pi /a,\pi /a,\pi /a)$ where $a$ is the cubic lattice
constant). A further complication is that at finite temperatures the minima
are sharp single peaks along (q00), sharp double peaks along (qq0), and
sharp triple peaks along (qqq) due to the thermal motions. Thus we displaced
our choice of q slightly from the commensurate q's to obtain the best
spectra. When $q$ is in the first Brillouin zone $S({\bf q},\omega )$ gives
to first order only the longitudinal excitations since the particle
positions enter only as $q\cdot r$ [{\it Kaneko and Ueda}, 1989], thus we do
not expect to see clear peaks for transverse excitations. The complexity of
our spectra may reflect the fact that we are looking primarily at
longitudinal excitations, which are affected greatly by electrostatic
depolarization effects in finite crystals.

\section{Results and Discussion}

\subsection{Melting Curve}

Results of our simulations showing the enthalpies and volumes of solid and
liquid at P = 0 are shown in fig.1. As a check on system-size effects, the
calculations were repeated using a 512-atom system. No system-size effects
were seen in volume or enthalpy. For any temperature, one may obtain the
difference in volume and enthalpy between solid and liquid, either by
interpolating or extrapolating. From the figure one can see that there is
some temperature variation in both of these quantities. However, because
both are increasing functions of temperature, their quotient is less
sensitive.

To obtain the melting curve, the melting temperature at each pressure was
estimated to be the temperature of the liquid just beyond the limit of
superheating of the solid. $\Delta $V and $\Delta $H of melting were then
calculated for those temperatures, and Eq. 1 numerically integrated to
obtain new estimates of the melting temperatures. This process was repeated
until the melting temperature converged. The resulting melting curve is
shown in fig. 2.

We show the fractional change in volume on melting and $\Delta
$H$_m$ versus pressure in fig. 3. Agreement is quite good for the
volume of melting with Cohen and Gong [{\it 1994}] and with Vocadlo
and Price [{\it 1996}]. The enthalpy of melting agrees within the
precision of the earlier cluster results of Cohen and Gong.  The
$\Delta $H$_m$ of Vocaldo and Price are in good agreement with our
results at zero pressure, but show a drop with increasing
pressure. We consider it more likely for $\Delta $H$_m$ to increase
with pressure as we find, since ionic interaction energies increase
with pressure and thus it should take more energy to diffuse atoms
or disorder the system on melting for ionic forces.

There is a large discrepancy between our predicted melting curve and
the experimental melting results of Zerr and Boehler, amounting to a
factor of three in the slope dT/dP (114 K/GPa versus 35,
respectively). This implies a discrepancy in $\Delta $V$_m$ and/or
$\Delta $H$_m$ through the Clapeyron equation
(eq. 1). Unfortunately, neither $\Delta $H$_m$ nor $\Delta $V$_m$
have been measured directly for any alkaline earth
oxide. Nevertheless, our value for $\Delta $H$_m$ is consistent with
literature estimates [{\it Chase,} 1985]. However, even in
principal we could not prove that the experimental result is
incorrect using a theoretical model; rather further experiments are
called for. One possible explanation of the origin of the
discrepancy would be Ar solubility in the MgO melt in the
experiments, thus depressing the melting curve at high
pressures. Other issue would be the melting criterion which perhaps
is more difficult at non-zero pressure. In any case finding the
origin of the discrepancy is important since similar methods are
being used to obtain melting curves for other materials as well.

\subsection{Power Spectrum}

The power spectrum $S(q,\omega )$ is quite complicated for our clusters and
is not fully understood. The complication over periodic boundary conditions
is that we observe not only phonon-like excitations, but also free
oscillations of the cluster [{\it Ozaki, Ichihashi, and Kondow}, 1991],
which are particularly evident at zero pressure due to the free surfaces.
Fig. 4 shows the power spectrum as a function of q at zero pressure. The
dispersive peaks are phonon-like modes and those whose frequencies vary
little with $q$ are the free oscillations. The latter are most obvious at
the smallest $q$ since at small $q$ the response is effectively averaged
over the whole cluster. Similar behavior was observed in tiny Ar$_{13}$ clusters [{\it Bhattacharya, Chen, and Mahanti}, 1996]. The phonon peaks
are also split due to the shape of the cluster and its small finite size.
Finally, as mentioned above, we expect to primarily see only longitudinal
motions as we have only considered $q$ in the first Brillouin zone. The
frequencies obtained using lattice dynamics in periodic boundary conditions
at the average volume of the cluster at zero pressure (21 \AA $^3$) are
indicated by the straight line segments. It is clear from the complicated
spectra and the complex relationship (or lack thereof) between the peaks in $%
S(q,\omega )$ and periodic boundary conditions that the cluster, at least at
zero pressure, has a significantly different mode structure than a periodic
crystal. This must be kept in mind when interpreting our results. Until a
similar study is done in periodic boundary conditions we must consider the
behavior of the power spectrum we obtain to indicate properties of clusters,
not necessarily bulk MgO.

In any case, we do observe interesting behavior in the cluster $S(q,\omega )$
on increasing temperature towards melting. These results are shown in fig.
5. The most obvious changes on increasing temperature are the reduction in
intensity of the phonon and free oscillation peaks, and significantly before
the melting transition growth is observed in the low frequency response.
Fig. 6 shows the low frequency part of the response (3-19 cm$^{-1}$) as a
function of temperature. The rapid rise below the melting transition is
evidence of premelting behavior which may be related to increases in heat
capacity on approach to melting observed in many systems [{\it Richet and
Fiquet}, 1991]. It may also be related to instabilities driven by
dislocations in larger systems [{\it Lund}, 1992].

Since our clusters have surfaces, the first obvious question is whether the
low frequency power we observe on approach to melting is related to surface
melting or other surface changes, or is in the bulk of the cluster. In order
to determine where the low frequency power is localized we filtered the
Fourier representation of the density function (eq. \ref{density}) and then
transformed back to real space. Fig. 7a shows a representative snapshot of
resulting low frequency, low q weight contribution on each atom, with the
radii of the atoms proportional to their contribution to the low frequency
response at 100 GPa and T= 11360 K. It is clear that the low frequency power
is a bulk effect and not localized at the surface.

Next we consider what dynamics is involved in the low frequency part of $%
S(q,\omega )$ by tracing out selected trajectories of atoms, and coloring
the trajectories by their contribution to the low frequency power (fig. 7b).
If all trajectories were depicted it is difficult to make anything out, but
it is clear from examining many trajectories and snapshots that at high
pressures the low frequency power is due to complicated many atom exchanges.

\section{Conclusions}

We have performed MD simulations of periodic bulk and finite 1000 atom
clusters of the melting of MgO. We obtain a melting curve that has a slope
three times greater than that obtained experimentally, but obtain reasonable
volumes and enthalpies of melting. The experiments should be repeated. If
indeed the ionic model is as far off from experiment as the current results
show, something exotic must occur in the electronic structure and bonding of
MgO liquid. The alternative is that there is some systematic error in the
experiment. We have also studied the power spectrum on heating in MgO
clusters and find evidence for premelting phenomena involving many atom
exchanges. It is possible that the onset of these many atom exchanges helps
lead to the melting instability, or put another way to the thermodynamic
destabilization of the crystal.

\section{Acknowledgments}

We thank Iris Inbar and Joe Feldman for helpful discussions.  Mark
Kluge performed the periodic bulk simulations. This work was
supported by NSF EAR-9418934, and computations were performed at the
Carnegie Institution of Washington and on the CM-5 at the National
Center for Supercomputing Applications.

\pagebreak
\section{Figure Captions}

Figure 1.\label{fig_v} Volumes and enthalpies of crystalline and liquid MgO
as functions of temperature at different pressures. The open diamonds and
triangles are for 512 atom supercells; other symbols are for 64 atom
supercells. The agreement for the different size periodic systems indicates
that we are converged for the volume and enthalpy. $\Delta V$ and $\Delta H$
are indicated at 0 GPa and 100 GPa respectively. (a) volume, (b) enthalpy.

Figure 2.\label{fig_melt} Melting curve for MgO. The present curve is
computed from the MD VIB volume and enthalpy data by integrating the
Clapeyron equation (eq. 1) starting at 3200 K at 0 GPa. The Cohen and Gong
[1994] curve was obtain by reversing the melting transition for clusters and
extrapolating to infinite system size using the PIB model. The Vocadlo and
Price [1995] curve was obtained by melting periodic systems with an
empirical potential. There is almost perfect agreement between the current
results and the Vocadlo and Price results. There is a factor of three
discrepancy in the slope with the experimental results of Zerr and Boehler.

Figure 3. Fractional (a)\ volume of melting and (b) enthalpy of melting
versus pressure.

Figure 4. $S(q,\omega )$ for P=0 as a function of ${\bf q}$. The
sharp low frequency peaks at 60 and 100 cm$^{-1}$ are resonances of
the cube. Acoustic phonons are visible at higher frequencies and are
dispersive. (a) $(q00)$ (b) $(qq0)$ (c) $(qqq)$.

Figure 5.  $S(q,\omega )$ for $q=(2 \pi/5a, 2 \pi/5a, 0)$ at (a) 0
GPa, (b) 100 GPa, and (c) 150 GPa.  The curves are offset for each
temperature shown.  At 0 GPa, the crystal has not yet melted at 3200
K, but the low frequency peak is evident.  At 100 GPa, melting
occurs between 11360 and 11600 for this cluster size, and a low
frequency peak arises before melting.  At 150 GPa melting occurs
between 12800 and 13300K, and there is a low frequency peak at 12800
K.

Figure 6. Low frequency power (integrated from 3-19 cm$^{-1}$) as a function
of temperature at different pressures. At the melting transition the low
frequency power seems to drop. Note that melting requires superheating at
zero pressure, but little hysteresis is observed at high pressures. Also,
note that melting occurs at higher temperatures than the bulk at high
pressures.

Figure 7. (a) Representation of selected atomic contributions to the
low frequency power at 100 GPa and T=11360 in the crystal before
melting. (b) Representation of trajectories of selected atoms, with
the shade chosen to show the contribution to the low frequency power
at P=100 GPa, and T= 11360 K. The black trajectories show atoms that
do not contribute to the low frequency response--they show only
oscillating, non-diffusive motions. Atoms that contribute to the low
frequencies response do diffuse and seem to exchange with other
atoms in complicated motions; the greater the low frequency
response, the lighter the shade of grey.

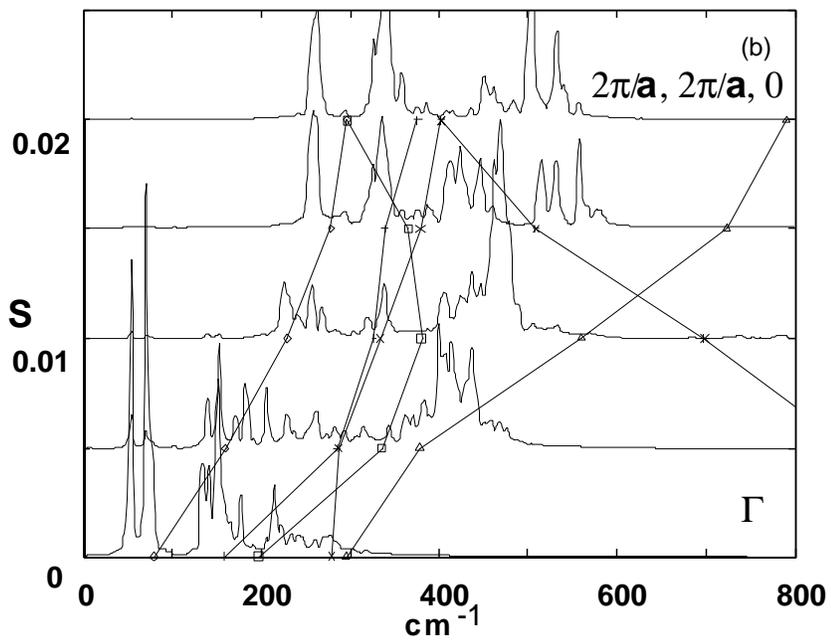
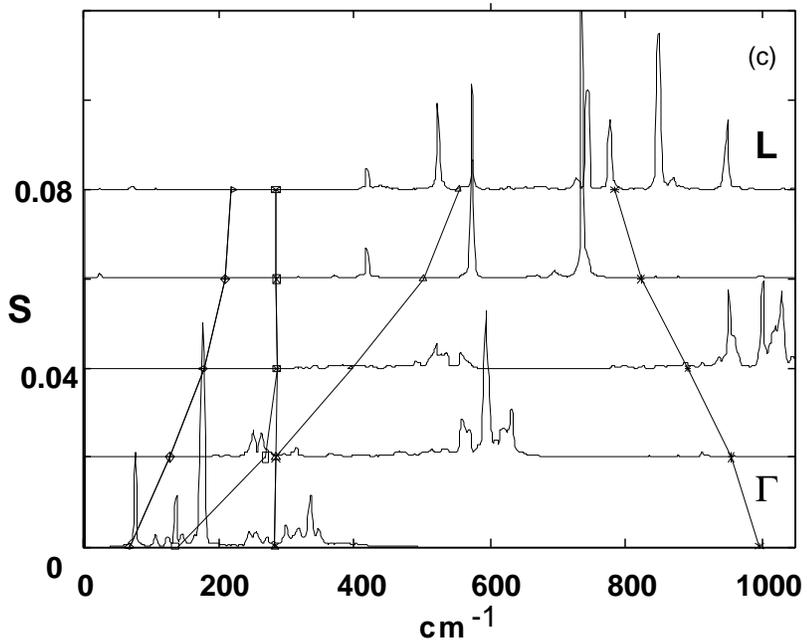
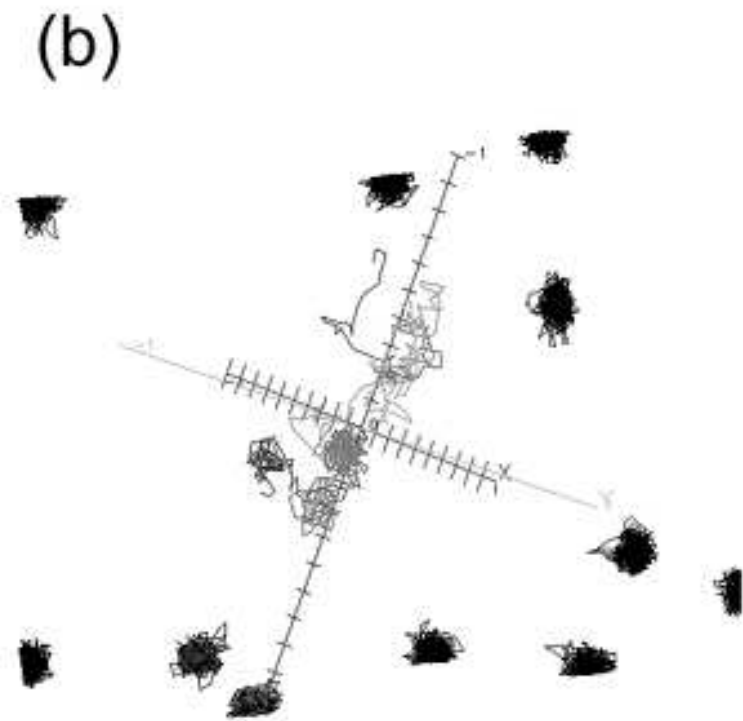
\end{document}